\documentclass[final]{ws-ijmpa}
\usepackage[super,compress]{cite}
\usepackage{graphicx}
\usepackage{hyperref}
\hypersetup{colorlinks,urlcolor=black,citecolor=black,linkcolor=black,filecolor=black}

\usepackage{xcolor}
\usepackage[normalem]{ulem}

\begin{document}
\markboth{Bergen pCT Collaboration}{Reconstruction of proton relative stopping power}

%
\catchline{}{}{}{}{}
%

\title{
 Reconstruction of proton relative stopping power with a granular calorimeter detector model
}

\author{
M. Aehle\textsuperscript{a}, 
J. Alme\textsuperscript{b}, 
G.G. Barnaföldi\textsuperscript{c},
G. Bíró\textsuperscript{c,l},
T. Bodova\textsuperscript{b}, 
V. Borshchov\textsuperscript{d}, 
A. van den Brink\textsuperscript{b}, 
M. Chaar\textsuperscript{b}, 
B. Dudás\textsuperscript{l}, 
V. Eikeland\textsuperscript{e}, 
G. Feofilov\textsuperscript{f}, 
C. Garth\textsuperscript{g}, 
N.R. Gauger\textsuperscript{a}, 
O. Grøttvik\textsuperscript{b}, 
H. Helstrup\textsuperscript{h}, 
S. Igolkin\textsuperscript{f}, 
Zs. Jólesz\textsuperscript{c,l},
R. Keidel\textsuperscript{i}, 
C. Kobdaj\textsuperscript{j},
T. Kortus\textsuperscript{a}, 
L. Kusch\textsuperscript{v}, 
V. Leonhardt\textsuperscript{g}, 
S. Mehendale\textsuperscript{b}, 
R. Ningappa Mulawade\textsuperscript{i}, 
O.H. Odland\textsuperscript{k, b}, 
G. O'Neill\textsuperscript{b}, 
G. Papp\textsuperscript{l}, 
T. Peitzmann\textsuperscript{e}, 
H.E.S. Pettersen\textsuperscript{k}, 
P. Piersimoni\textsuperscript{b,m}, 
M. Protsenko\textsuperscript{d}, 
M. Rauch\textsuperscript{b}, 
A. Ur Rehman\textsuperscript{b}, 
M. Richter\textsuperscript{n}, 
D. Röhrich\textsuperscript{b}, 
J. Santana\textsuperscript{i}, 
A. Schilling\textsuperscript{a}, 
J. Seco\textsuperscript{o, p}, 
A. Songmoolnak\textsuperscript{b, j}, 
J. Rambo Sølie\textsuperscript{q}, 
G. Tambave\textsuperscript{w}, 
I. Tymchuk\textsuperscript{d}, 
K. Ullaland\textsuperscript{b}, 
M. Varga-Kőfaragó\textsuperscript{c}, 
L. Volz\textsuperscript{s}, 
B. Wagner\textsuperscript{b}, 
S. Wendzel\textsuperscript{i}, 
A. Wiebel\textsuperscript{i}, 
R. Xiao\textsuperscript{b, t}, 
S. Yang\textsuperscript{b}, 
H. Yokoyama\textsuperscript{e}, 
S. Zillien\textsuperscript{i} on behalf of the Bergen pCT collaboration
}

\address{
a) Chair for Scientific Computing, University of Kaiserslautern-Landau, 67663 Kaiserslautern, Germany; \\
b) Department of Physics and Technology, University of Bergen, 5007 Bergen, Norway; \\
c) HUN-REN Wigner Research Centre for Physics, 29--33 Konkoly--Thege Mikl\'os \'ut,\\ H-1121 Budapest, Hungary; \\
d) Research and Production Enterprise "LTU" (RPELTU), Kharkiv, Ukraine; \\
e) Institute for Subatomic Physics, Utrecht University/Nikhef, Utrecht, Netherlands; \\
f) St. Petersburg University, St. Petersburg, Russia; \\
g) Scientific Visualization Lab, University of Kaiserslautern-Landau, 67663 Kaiserslautern, Germany; \\
h) Department of Computer Science, Electrical Engineering and Mathematical Sciences, Western Norway University of Applied Sciences, 5020 Bergen, Norway; \\
i) Center for Technology and Transfer (ZTT), University of Applied Sciences Worms, Worms, Germany; \\
j) Institute of Science, Suranaree University of Technology, Nakhon Ratchasima, Thailand; \\
k) Department of Oncology and Medical Physics, Haukeland University Hospital, 5021 Bergen, Norway; \\
l) Institute for Physics, Eötvös Loránd University, 1/A Pázmány P. Sétány, H-1117 Budapest, Hungary; \\
m) UniCamillus - Saint Camillus International University of Health Sciences, Rome, Italy; \\
n) Department of Physics, University of Oslo, 0371 Oslo, Norway; \\
o) Department of Biomedical Physics in Radiation Oncology, DKFZ—German Cancer Research Center, Heidelberg, Germany; \\
p) Department of Physics and Astronomy, Heidelberg University, Heidelberg, Germany; \\
q) Department of Diagnostic Physics, Division of Radiology and Nuclear Medicine, Oslo University Hospital, Oslo, Norway; \\
r) Budapest University of Technology and Economics, Budapest, Hungary; \\
s) Biophysics, GSI Helmholtz Center for Heavy Ion Research GmbH, Darmstadt, Germany; \\
t) College of Mechanical \& Power Engineering, China Three Gorges University, Yichang, People's Republic of China;\\
u) Department of Radiology, Faculty of Medicine, Chulalongkorn University, 1873 Rama IV Rd, Pathum Wan, Bangkok, 10330, Thailand;\\
v) Eindhoven University of Technology, Eindhoven, Netherlands;\\
w) Center for Medical and Radiation Physics (CMRP), NISER Bhubaneswar 752050, India.
}

\maketitle


\begin{abstract}
Proton computed tomography (pCT) aims to facilitate precise dose planning for hadron therapy, a promising 
and effective method for cancer treatment. Hadron therapy utilizes protons and heavy ions to deliver well focused doses of radiation, leveraging the Bragg peak phenomenon to target tumors while sparing healthy tissues. The Bergen pCT Collaboration aims to develop a novel pCT scanner, and accompanying reconstruction algorithms to overcome current limitations.
This paper focuses on advancing the track- and image reconstruction algorithms, thereby enhancing the precision of the dose planning and reducing side effects of hadron therapy. A neural network aided track reconstruction method is presented.
\end{abstract}

\section{Introduction}	
\label{sec:intro}

Hadron therapy~\cite{wilson1946radiological}, an advanced form of radiation therapy, employs protons and heavy ions like helium, carbon and oxygen to treat cancer, offering significant advantages over traditional X-ray therapy~\cite{protonBerkely, mayles2007handbook, pCTReview2018}. Unlike X-rays, which deposit energy throughout their path, hadron therapy leverages the Bragg peak~\cite{Braggpaper} of charged particles to deliver the majority of radiation energy precisely at the tumor site, minimizing damage to surrounding healthy tissues~\cite{fundCompTomo}. Proton therapy, the most widely used form of hadron therapy of today, benefits from its ability to finely control the depth of energy deposition, enhancing treatment accuracy. Proton computed tomography (pCT)~\cite{pctarticle} further augments this precision by using protons for both imaging and treatment, directly determining 
electron densities of the tissue  and enabling more accurate dose planning. This approach reduces uncertainties associated with X-ray CT conversions, thereby improving the efficacy and safety of cancer treatments~\cite{Durante2021}.

Specialized track- and image reconstruction are crucial for pCT due to the unique behavior of massive charged particles compared to X-rays~\cite{wang2010use, Rambo_S_lie_2020, pctdetector, alme2020high, ordonez2019fast, solie2020image}. Charged particles like protons can undergo significant scattering within the patient's body, making it challenging to determine their exact paths without specialized reconstruction algorithms~\cite{lynch1991, williams2004most, schulte2008maximum, krah2018comprehensive}. Traditional X-ray methods cannot account for these interactions, leading to inaccuracies of the relative stopping power (RSP: the stopping power of the given material compared to water) of the protons. Image reconstruction techniques, such as the Richardson\,--\,Lucy algorithm, are tailored to handle the specific requirements of proton data (such as the most likely paths inside the phantom/patient) in an iterative way~\cite{brooks1975theory, penfold2015techniques, gordon1970algebraic, gilbert1972iterative, richardson1972bayesian, ingaramo2014richardson, lucy1974iterative, Sudar:2022sww}, while advanced track reconstruction algorithms ensure precise determination of the necessary proton energies and directions after the phantom. These specialized methods are essential to achieve the high spatial resolution, accuracy for the RSP, and efficient computation necessary for effective and clinically viable pCT.

In this work we present two crucial software aspects of a viable pCT infrastructure
~\cite{solie2020monte, alme2020high, solie2020image}, namely the track reconstruction algorithms based on the detector signals, and the evaluation of the image reconstruction algorithms. The paper is structured as follows: in Section~\ref{sec:intro2} the aims of the study are introduced; in Section~\ref{sec:track} we give details about the track reconstruction for simulated data, and in Section~\ref{sec:image} the results of the RSP reconstruction are shown, also for simulated data. In Section~\ref{sec:summary} we summarize the results.

\section{Aims of the pCT development}
\label{sec:intro2}

The Bergen proton CT Collaboration aims to develop a state-of-the-art proton computed tomography scanner that utilizes advanced particle tracking capabilities with the ALPIDE monolithic active pixel sensor technology---in the followings the up-to-date setup is briefly summarized.
~\cite{AliceAlpide, pctdetector, solie2020monte, alme2020high, solie2020image, Aehle_2023} 

The developed detector system follows a  single-sided scanner design, omitting the upstream tracker between the beam and the phantom/patient. The downstream detector employs list-mode data acquisition with a granular tracking calorimeter design, utilizing 2 tracker layers without absorbers (for the direction measurement of the incoming particles), separated by 57.8 mm, followed by 41 calorimetric layers, where an additional, 3.5 mm thick aluminum absorber is attached to each tracking layer, the later are separated by 5.5 mm. Each tracking layer consists of $5.4\times 10^7$ pixel ALPIDE\cite{AliceAlpide} sensors.
The aim of this design is to reconstruct the tracks (defined as the kinetic energy and direction) of every single incoming particle, which is then utilized to reconstruct the relative stopping power map of the phantom.

The collaboration involves multiple international institutions, focusing on designing and assembling the mechanical and electrical components, conducting Monte Carlo simulations, data analysis, and hardware testing, including beam tests of the detector prototypes~\cite{Aehle_2023}. In this work we focus on the development of particle track reconstruction based on the Monte Carlo simulation of detector signals, which is then followed by an iterative image reconstruction method to reconstruct the proton RSP from the synthetic data.

\section{Track reconstruction}
\label{sec:track}

\begin{figure}[h!]
\begin{center}
    \includegraphics[width=0.8\textwidth]{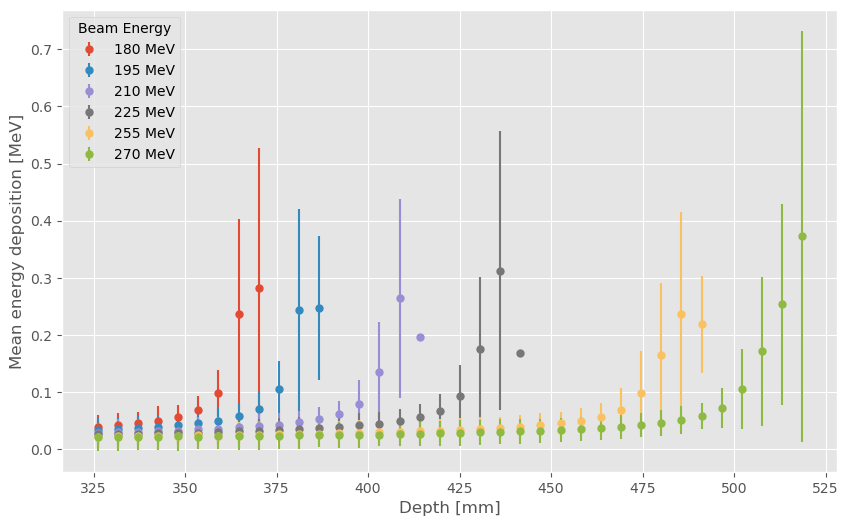}
    \caption{ The mean deposited energy of protons as a function kinetic energies in the pCT detector system with a 160~mm thick water phantom.
    }
    \label{fig:Bragg_peak}
\end{center}
\end{figure}

During the treatment of the patient, the energy of the beam is tuned in a way that the Bragg peak is well localized inside the tumor---this tuning is illustrated on Fig.~\ref{fig:Bragg_peak}, where the position of the peak of the energy deposition inside a water phantom depends on the energy of the proton beam. However, during the dose planning and tomography mode, the maximum of the deposited energy is well behind the patient, inside the detector.

In Fig.~\ref{fig:Bragg_peak} we show {\tt GATE} simulated deposited energy {\it averaged} over $10^5$ protons, at different initial energies. Due to fluctuation in the beam position and direction, and the scattering itself being a stochastic process the "Bragg peak" is smoothened with hits even above it~\cite{Rambo_S_lie_2020}.

In order to reconstruct the RSP map of the patient, a precise measurement of the protons that traversed the patient is necessary 
~\cite{AliceAlpide, pctdetector}.
In the current work, using the synthetic data a well determined simulation environment is used: 
during a single readout frame, as low as $\mathcal{O}(100)$ proton trajectories are measured. On the other hand, for a full scan $\mathcal{O}(10^6 - 10^8)$ trajectories (measured from several readout frames of variable sizes)
are needed~\cite{pctdetector}---therefore, it is essential to accurately reconstruct the tracks from the raw pixel signals within a medically reasonable time. 

The Kalman filter is a widely used approach in high-energy physics for track parameter estimation---however, it is traditionally computationally expensive~\cite{KalmanFilter}. 
In contrast to that, the reconstruction of the trajectories from the pixel clusters (referred to as \textit{hits}) can be seen as a matching problem, finding the proper connections of hits between the subsequent layers.
One can represent this problem as a bipartite graph matching \cite{bip_karman} and use the "Hungarian algorithm", described by Kuhn~\cite{hungarian_alg}.  An optimized and more recent variant of this method is the Ford\,--\,Fulkerson method~\cite{Ford_Fulkerson}, which is able to provide more accurate results at the cost of increased computational time. 
Deep neural networks are also used in some cases to enhance or replace traditional particle track reconstruction algorithms~\cite{hep_Dl,DBLP:journals/corr/abs-1812-03859}.

Due to the occasional large-angle, inelastic scattering of the protons and overlapping hits, the ratio of good matches is decreasing by 2-3\% after each detector layer. Following to these consecutive losses, approximately 70\% of the initial protons reach the typical position of the average Bragg peak (around layer numbers 20-24), and only 10\% of the initial proton number reaches beyond the peak with further 1-3 layers. This is due to the stochastic nature of the energy loss and due to the geometry modeling of the pencil beam, with a divergence (defined by the standard deviation of the normal probability density function) of 2.5 mrad and spot size (defined by the standard deviation of the normal probability density function in x and y directions) of 2 mm~\cite{Rambo_S_lie_2020}. We note, that the 'average' Bragg peak is easily defined for real experimental setup as well.

\begin{figure}[h!]
\begin{center}
    \includegraphics[width=0.8\textwidth]{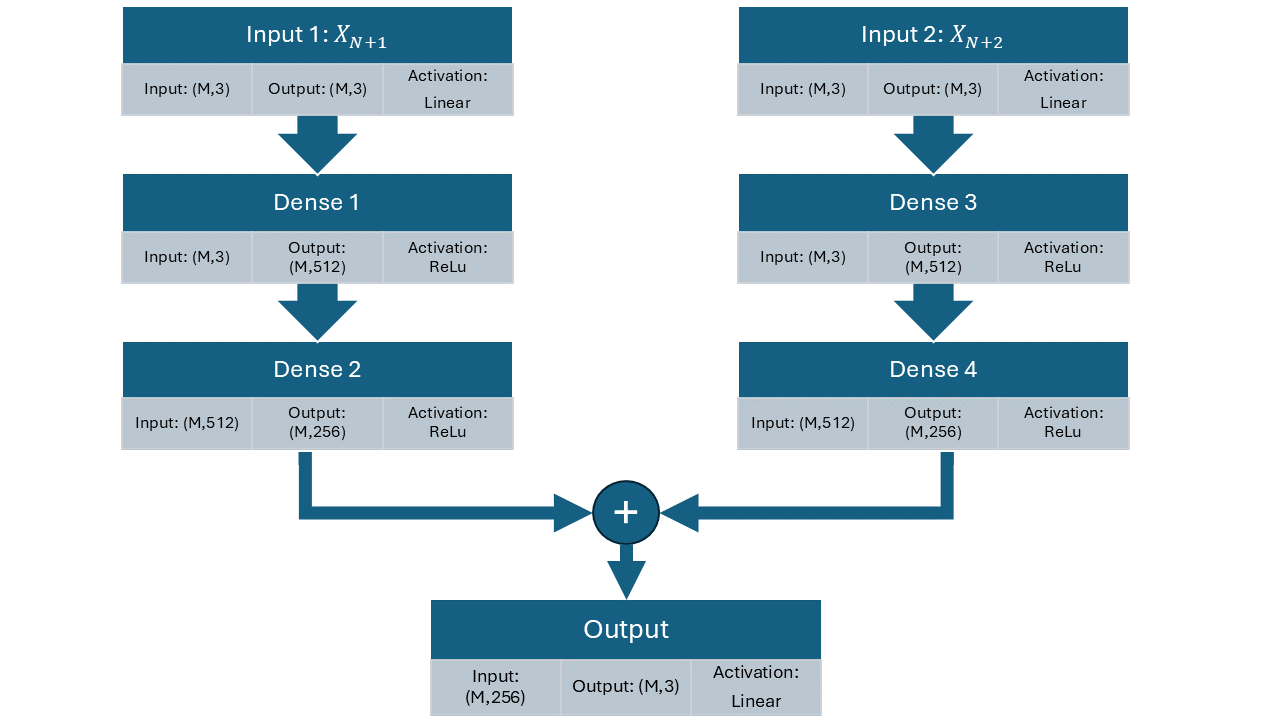}
    \caption{The structure of the position prediction model. }
    \label{fig:model_struct}
\end{center}
\end{figure}

Here, the matching problem is addressed by the Sinkhorn algorithm~\cite{Sinkhorn, Sinkhorn2}, in which the connection probability matrix between the detector hits is created. 
This method can give us good results for matching and also can be implemented with deep learning frameworks. This allows us not just compatibility with other GPU based algorithms that we use,   
but also makes runtime of the overall reconstruction more manageable than any other algorithm we applied, such as the Hungarian algorithm or the Ford\,--\,Fullkerson method. 
This is achieved by maximizing the following Sinkhorn operator:
\begin{equation}
\label{eq:sinkhorn}
    S(X_N,X_{N+1}) = \exp\left[-\frac{||X_{N}-X_{N+1}||}{T}\right],
\end{equation}
where $||X_{N}-X_{N+1}||$ is the euclidean distance matrix of the hits to be matched in two consecutive layers, $N+1$ and $N$ (note, that we are evaluating the track starting from its last hit, in reverse order), and $T$ is a parameter to be optimized. The vector $X_N$, contains all the hits of the layer, with each entry being a vector itself with the 2 positional coordinates of the hit ($x, y$) and the deposited energy in the layer, forming a 3-vector, and Eq.~\eqref{eq:sinkhorn} defines a matrix connecting the $i$\textsuperscript{th} hit of layer $N$ with the $j$\textsuperscript{th} hit of layer $N+1$. Note that the number of hits may differ in the two layers, hence the Sinkhorn algorithm automatically handles particles that leave the detector system. Next, the rows and columns of $S$ are normalized alternately for a given number of iterations, in order to achieve a probability matrix such that
\begin{equation}
\label{eq:sinkhorn2}
    \sum\limits_y S_{x,y} = 1.
\end{equation}
Alternatively, to improve matching efficiency (defined as the ratio of \textit{good matches}/\textit{all matches}), a fully connected deep learning model is implemented. 
The structure of the model can be seen in Fig.~\ref{fig:model_struct}, with $M=100$ being the maximal number of protons in a readout frame.

In this enhanced version, the hit positions of layer $N$ is predicted from the preceding two layers (in the detector geometry, going from backward to forward): 
\begin{equation}
    Y_{N} = \mathrm{model}_N(X_{N+1},X_{N+2}) \,,
\end{equation}
and in the Sinkhorn operator~\eqref{eq:sinkhorn} we substitute $X_{N+1}$ with $Y_N$. 
We used separately trained models for each layer, with the same structure as shown in Fig.~\ref{fig:model_struct}. In the case of the calorimetric layers, by applying the same model for each layer does not change the results. However, when tracking layers (without the absorbers) are also involved, due to the change of geometry, separate models were needed.
Thereafter, the predicted hits are matched with the measured hits of the \textit{same layer}, resembling to the Kalman filter approach. This method works especially well for the tracker layers, where otherwise proton trajectories are more difficult to distinguish due to the larger physical distance.

For training and evaluation of the track reconstruction algorithm, the {\tt GATE}\cite{Gatetoolkit, g41, g42,g43} medical simulation software is utilized, following the realistic beam model of the Bergen pCT Collaboration~\cite{Rambo_S_lie_2020}. For evaluation we selected all the primary protons within the simulation.

\begin{figure}[h!]
\begin{center}
    \includegraphics[width=0.75\textwidth]{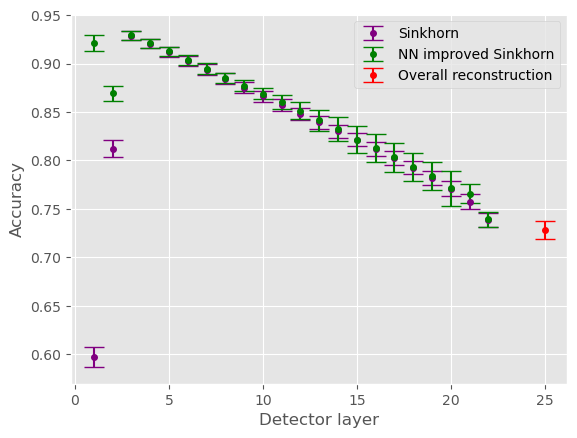}
    \caption{Particle hit matching accuracy between layers using the Sinkhorn algorithm (purple markers) and including our approach with neural network (NN) improved Sinkhorn model (green markers). Overall reconstruction (red marker) shows the accuracy of fully reconstructed tracks using both neural network and the consistency criterion. A 160~mm long water phantom was used in the simulation.}
    \label{fig:matching_accuracy}
\end{center}
\end{figure}
Figure~\ref{fig:matching_accuracy} shows the matching accuracy (the ratio of correctly matched hits to all matched hits, in percentage) of the algorithms. The NN-enhanced version significantly improves accuracy in the first two detector layers. In the Bergen pCT setup, they are tracking layers, with no absorber and much further separated, than the subsequent calorimetric layers.  

It is an important remark that for medical applications only a correctly reconstructed track can be accepted. A track is called \textit{perfectly matched} if every detector hit is correctly assigned to the corresponding track during the reconstruction -- any mismatch generates greater uncertainty in the determination of the energy. 

The cumulative precision of the track reconstruction, simply using the Sinkhorn algorithm is $48.7\% \pm 0.5\%$, i.e. the true tracks are less than 50\% of all identified tracks.

To further increase the accuracy of matching we dropped all the tracks, where the Sinkhorn matching produces different result row-wise compared to column-wise (consistency criterion). Doing so for the {\em whole} tracks still 70\% of the initial tracks are present.
Combining the NN refinement with the consistency criterion we managed to increase the precision to $72.8\% \pm 0.9\%$
(the NN refinement alone brings the precision to 53\%, while the consistency alone to 58\%).
The whole method is programmed in TensorFlow~\cite{tensorflow}, and can be run on GPUs.

As a last step of the track reconstruction, for imaging we need to determine the $E_{phantom}$ kinetic energies of protons after the phantom (before the detector).
To fit $E_{phantom}$ we used simulated data with water phantom thicknesses from 100 mm to 200 mm in 20 mm steps. We found, that $E_{phantom}$ correlates the most with the position of Bragg peak, $N$, the layer of the maximal energy deposit. To get rid of the nuclear interactions, we filtered the data to be near the ensemble averaged Bragg peak (within $\pm 1$ layer, which corresponds to roughly $\pm8$ MeV spread in the initial energy in our setup).

In real experiment the average Bragg curve may be extracted from all identified tracks at a given setup (energy, angle, position). Since most of the collisions are small angle Coulomb scattering, they dominate the average---as a consequence it is possible to filter out the inelastic scattered protons that deviate from the average~\cite{solie2020image}.

The fine tuning is done with the use of the deposited energies in the adjacent layers, normalized to the Bragg peak layer's one:
\begin{equation}
    \label{eq:pnnv}
    R_{\pm} =\frac{E_{N\pm 1}}{E_{N}} \,, 
\end{equation}
where $N$ is the layer number, where the Bragg peak is positioned, $E$ is the deposited energy of the corresponding detector layer. 
The energy is thus calculated as: 
\begin{equation}
  \label{eq:en_fit}
    E_{phantom} = \alpha \cdot N + E_0 + \gamma_{-} R_{-}  + \gamma_{+} R_{+} \,,
\end{equation}
where the fitted parameters are summarized in Table~\ref{tab:params_tab}.

\begin{figure}[h!]
    \centering
    \includegraphics[width=.49\textwidth]{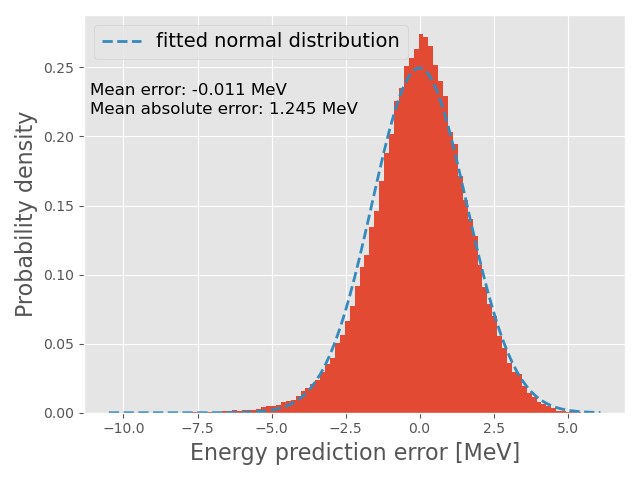}
    \includegraphics[width=.49\textwidth]{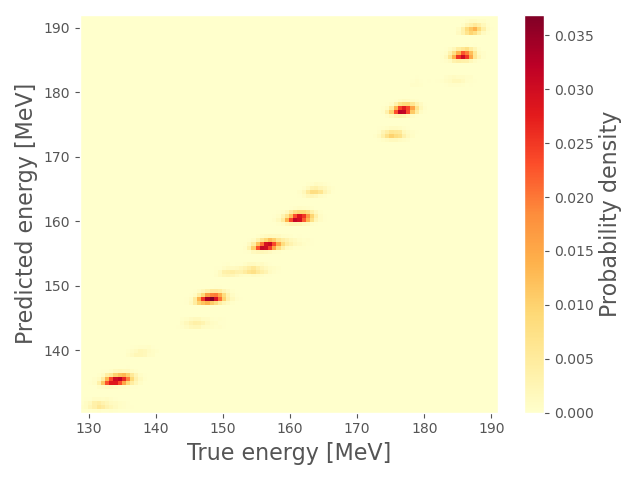}
    \caption{The distribution of the error in energy prediction (left panel) and the 2 dimensional histogram of the true values and the predicted values, calculated from Eq.~\eqref{eq:en_fit}  (right panel).
    }
    \label{fig:ind_losses}
\end{figure}

\begin{table}[!ht]
\tbl{Parameters of the fit to the energy after the phantom.}
{
    \begin{tabular}{lcccc}
    \toprule
     Parameter & $\alpha$ (MeV)& $E_0$ (MeV) & $\gamma_{-}$  (MeV) & $\gamma_{+}$ (MeV)\\
    \colrule
    Values & 4.191 & 66.705 & 2.943 & 5.243\\
    \botrule
    \end{tabular}
     \label{tab:params_tab}
}
\end{table}
In this way we achieved uncertainty of $\sigma = 1.245$ MeV at 230 MeV beam energy. As on the left panel of Fig.~\ref{fig:ind_losses}  is shown, the difference between the true and the predicted values is centered around zero and is close to a Gaussian. On the right panel of Fig.~\ref{fig:ind_losses} the true values versus the predicted ones are shown. The cluster pattern corresponds to the finite set of the length of the water targets, used in generating the training set for the fit. 

Next, using the energy uncertainty, obtained above, we study the reconstruction accuracy of the RSP.
    
\section{Relative stopping power map reconstruction}
\label{sec:image}

The relative stopping power map (simply referred to as \textit{image}) of a given object is the proton-CT equivalent of an X-ray image, representing the characteristic energy loss of the protons inside the object~\cite{penfold2015techniques}. By measuring each proton trajectory that traverses the object, we can reconstruct the images by solving the following linear equation system:
\begin{equation} \label{eq:1}
    \boldsymbol{A}\cdot \boldsymbol{x} = \boldsymbol{y},
\end{equation}
where $y_i = A_{ij} x_j$ represents the measured, integrated  RSPs along the proton's trajectory, corresponding to the water-equivalent thickness seen by the proton, $\boldsymbol{x}$ is a $J$-dimensional vector containing the unknown RSP values in a given voxel, and the $\boldsymbol{A}_{I\times J}$ matrix contains the interactions between protons and the volume elements (for simplicity referred as 2D pixels): the matrix element $A_{ij}$ represents the path length of the $i$\textsuperscript{th} proton in the $j$\textsuperscript{th} pixel. Given that the elements of $y$ are known from energy loss measurements (based on the methods described in the previous section), the goal of image reconstruction is to determine $x$.

For image reconstruction with protons, techniques based on integral transformations (such as back projection and filtered back projection) are unsuitable because they assume no scattering, a condition that does not hold for protons. Consequently, the alternative class of reconstruction techniques, specifically the iterative algorithms, and more precisely, the statistical iterative algorithms, are more appropriate. Although these methods are not yet widely applied in CT image reconstruction, they are promising for this problem as they can accurately model energy loss. 

A notable type of iterative algorithm is the Richardson\,--\,Lucy algorithm, which approaches the problem with the so-called Maximum Likelihood - Expectation Maximization (ML-EM) method~\cite{lucy1974iterative, richardson1972bayesian}. 
This algorithm is an effective method for reconstructing images that are modified by blurring, which can be described by a point spread function, and results in a set of indistinguishable pixels. However, it has not yet been applied in medical imaging. The algorithm's solution is governed by the following equation:
\begin{equation} \label{eq:2}
    x_i^{k+1} = x_i^k \frac{1}{\sum_j A_{ij}} \sum_j \frac{y_j}{\sum_l A_{lj} x_l^k} A_{ij} \,.
\end{equation}

While this problem is relatively straightforward from a mathematical perspective, it is computationally expensive due to the necessity of processing millions of proton trajectories for an accurate RSP image. In this current study, the first, preliminary RSP result comparisons of our CUDA based implementation is shown. The input data for the algorithm was generated by {\tt GATE}~\cite{Gatetoolkit, jan2004gate}, assuming a quasi-ideal detector with perfect spatial resolution but with an estimated energy uncertainty of $\sigma=2$ MeV, where to be on the safe side, we have chosen a slightly higher value as determined in Section~\ref{fig:ind_losses}, and compared to the ideal case. 

\begin{figure}[h!]
    \centering
    \includegraphics[width=0.45\textwidth]{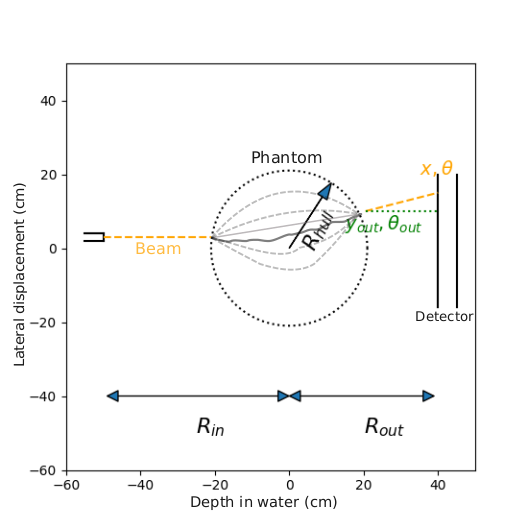}
    \caption{The schematic figure of MLP calculation (proton track - bold line, result of MLP calculation - solid line, added errors - dashed lines).}
    \label{fig:mlp}
\end{figure}

The final resolution of proton CT imaging is constrained by a number of factors, including the random scattering of protons upon entering the medium,  the various parameters of the imaging system and the proton beam~\cite{krah2018comprehensive}. Due to the fact that protons may suffer a sequence of Coulomb scatterings inside the patient, a major issue is that the elements of $\boldsymbol{A}$ can be determined only through computationally heavy, so called \textit{most likely path} (MLP) calculations, as it is depicted in Fig.~\ref{fig:mlp}.

In order to determine the MLP, it is necessary to measure the position and angle of the protons at distances $z = -R_{in}$ and $z = R_{out}$ from the centre of the phantom. These angles and distances can be obtained from the previously described track reconstruction, however, in the current work we use simply simulated data, to study what is the best performance of the image reconstruction model one can achieve. On the assumption that the proton travels in a straight line through the air, the measured values are then projected onto the outline of the phantom. Let us denote the projected transverse position and angle values by $y_0$ and $y'_0$, where the latter is regarded as a quantity describing the slope (i.e., the tangent of the angle) rather than an actual angle. From that on the intersection of the path of the proton and the outline of the phantom can be calculated, followed by the calculation of the MLP at the intersections. Various mathematical models that describe the multiple Coulomb scattering of protons can be employed to ascertain the most likely path of the proton~\cite{schulte2008maximum, williams2004most, krah2018comprehensive}. In our study, we employed a Cubic Spline Path calculation, which gives a reasonably good approximation of the MLP with manageable computation times~\cite{solie2020monte, fekete2015developing}. 

\begin{figure}[h!]
    \centering    
    \raisebox{0.37\height}{\includegraphics[width=0.37\textwidth]{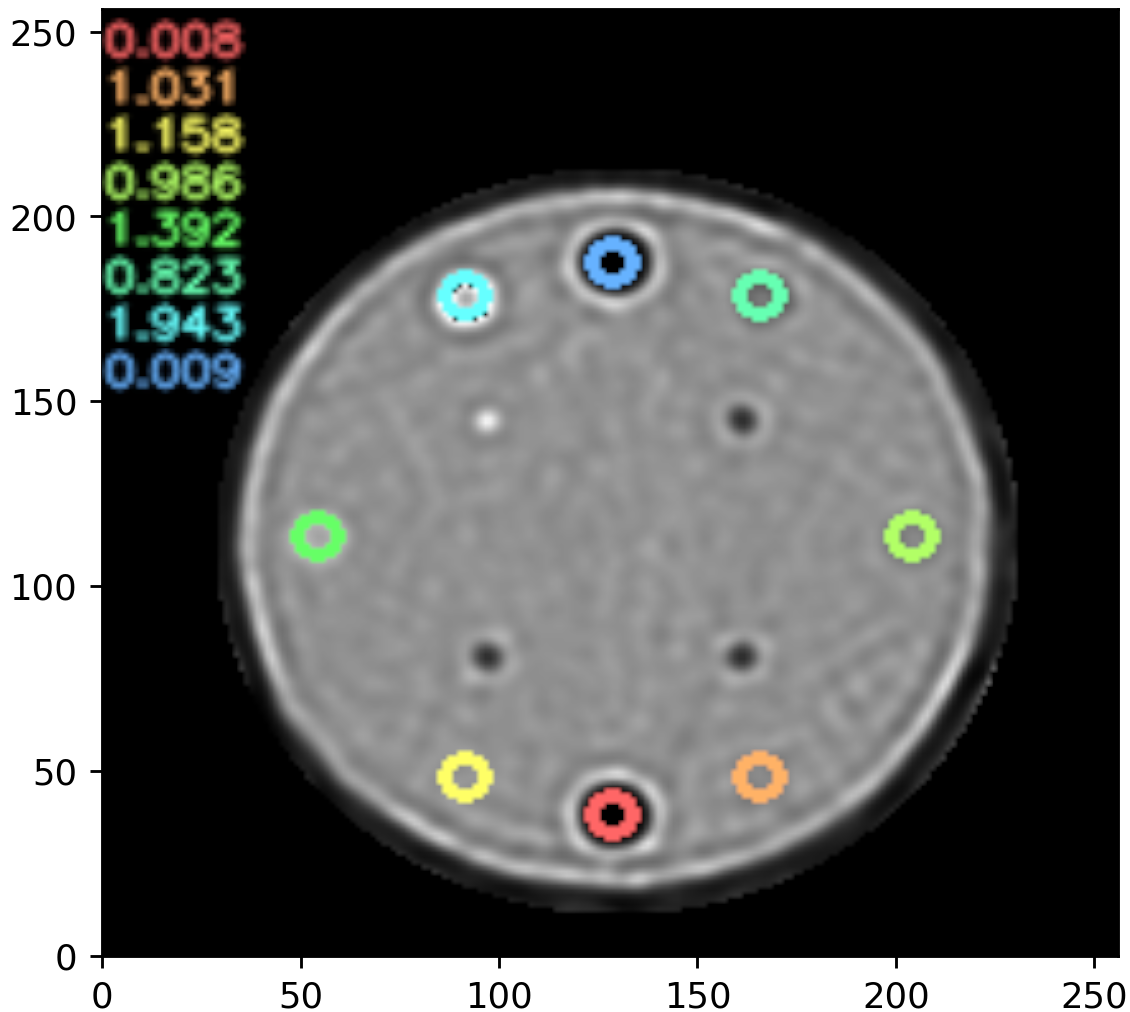}}
    \includegraphics[width=0.62\textwidth]{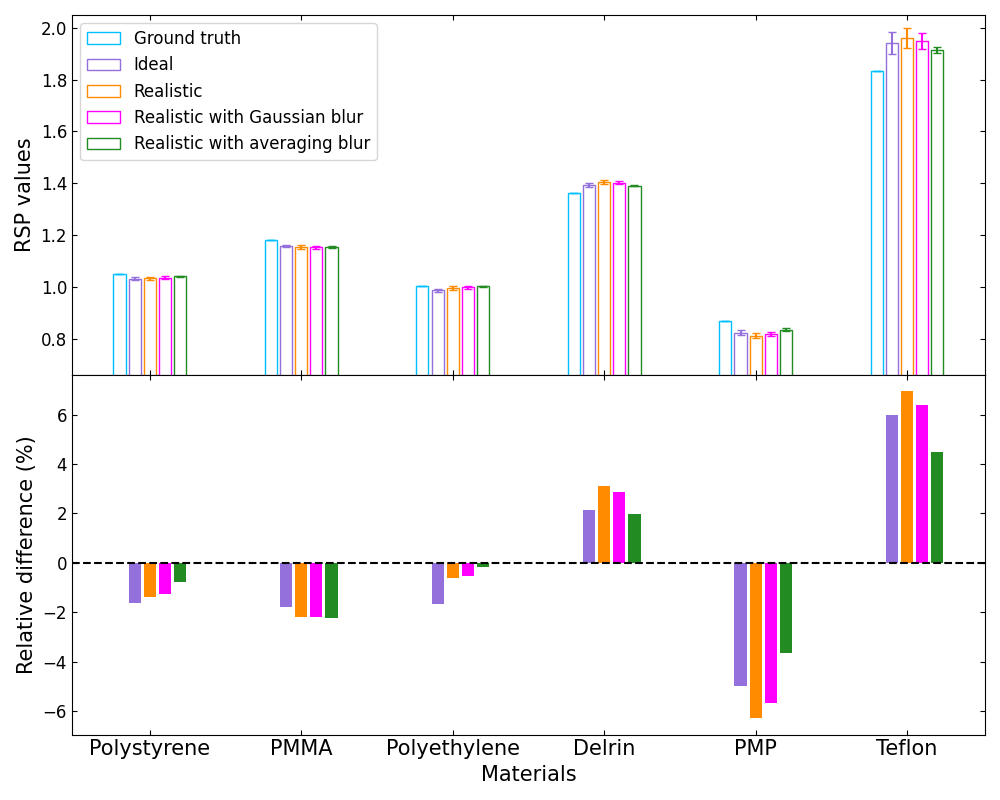}
    \caption{The reconstructed CTP404 phantom (left) and the relative differences between the reconstructed and the ground truth values (right).}
    \label{fig:ctp}
\end{figure}

The results of the RSP reconstructions are illustrated in Fig.~\ref{fig:ctp}, showing the CTP404 phantom~\cite{ctp404}, specifically designed for the density evaluation of imaging systems. It is a 150 mm diameter epoxy cylinder containing 8 inserts (each 12.2 mm in diameter) made of different materials (represented by different grayscale values and highlighted with different colors for better visibility), making it suitable for evaluating the reconstruction of proton RSP values. The reconstructed phantom image with the different RSP values can be seen on the left (with a 1 mm/pixel reconstructed resolution), while the plot on the right illustrates the relative difference between compared to the ground truth RSP values for the following cases:
\begin{enumerate}
    \item the \textit{ideal} reconstruction has no uncertainty in the proton energy;
    \item in the \textit{realistic} case, the uncertainty of the proton energy follows a Gaussian distribution with $\sigma = 2$ MeV;
    \item the \textit{Gaussian blur} and \textit{average blur} cases show the results where an additional blurring on RSP values is applied to the \textit{realistic} case in order to study the effects of the energy uncertainty. 
\end{enumerate}

As it is expected, the uncertainty in the proton energy is introducing variability in the reconstructed RSP values, resulting in a higher relative difference from the ground truth compared to the ideal case (6.9\% compared to 5.9\%). It results in a more noisy RSP distribution, which reflects the spread of information that the uncertainty brings, causing a slightly less precise reconstruction.

Any applied additional blurring mitigates the noise or other artifacts caused by the energy uncertainty, acting as a regularization: the Gaussian blurring has the effect of distributing pixel-wise RSP values based on a normal distribution, smoothing out high-frequency noise introduced by the proton energy uncertainty. As the \textit{right panel} of Fig.~\ref{fig:ctp} shows, where the \textit{realistic} case look more similar to the \textit{ideal} (with $\sim 6.4\%$ maximal difference), the blurring reduces the artifacts caused by uncertainty, which suggests that the uncertainty manifests as high-frequency noise or sharp variations.

On the other hand, the average blurring reduces sharp local variations more aggressively, which is reflected in even less relative difference values of $\sim 4.5\%$. The stronger regularization shows that this current RSP reconstruction method is sensitive to the noise that the energy uncertainty introduces. 

It is important to note, that although these blurring processes have the ability to mitigate the noise, and therefore possibly improving the RSP resolution of the result, it might cause unwanted effects in the contrast (i.e. the spatial resolution) of the reconstructed image---however, this is out of the scope of the current study.

\section{Summary}
\label{sec:summary}
We have developed a particle trajectory reconstruction method for the Bergen pCT Collaboration that can sufficiently match detector hits of protons in our detector system with a neural-network aided algorithm. This method is the combination of Neural Network and Sinkhorn matching. Our preliminary studies with simple fitting indicate the energy resolution to be better than 2 MeV, however, more sophisticated models are under development.

For image reconstruction with protons, traditional methods like back-projection are unsuitable due to proton scattering, so statistical iterative algorithms such as the Richardson-Lucy method are employed. A key challenge is calculating the most likely path (MLP) of protons, which involves complex modeling of multiple Coulomb scatterings. In this study we used cubic spline calculations for MLP, balancing accuracy and computational efficiency. Tests on the CTP404 phantom, containing different materials, showed less than 6\% discrepancy between the reconstructed and the ground truth RSP values. Our tests confirmed that the reconstruction is robust against energy uncertainties of up to 2 MeV. This suggests that this approach can be reliably applied in contexts where energy resolution fluctuations within this range are present.

\section*{Acknowledgments}

This work has been supported by the NKFIH grants OTKA K135515, as well as by the 2024-1.2.5-TÉT-2024-00022, 2021-4.1.2-NEMZ\_KI-2024-00031 and 2021-4.1.2-NEMZ\_KI-2024-00033 projects. The project is supported by the NKFIH DKOP-23 Doctoral Excellence Program of the Ministry for Culture and Innovation.
The authors acknowledge the research infrastructure provided by the Hungarian Research Network (HUN-REN) and the Wigner Scientific Computing Laboratory. Authors G. B., G. P. and B. D. were supported by the European Union project RRF-2.3.1-21-2022-00004 within the framework of the Artificial Intelligence National Laboratory.




\bibliographystyle{ws-ijmpa}
\bibliography{pCT_BP}

\end{document}